\documentclass[sigconf]{acmart}

\newcommand{\tcode}[1]{\texttt{\footnotesize{#1}}}

\usepackage{soul}

\AtBeginDocument{%
  \providecommand\BibTeX{{%
    \normalfont B\kern-0.5em{\scshape i\kern-0.25em b}\kern-0.8em\TeX}}}
    
\copyrightyear{2022}
\acmYear{2022}
\setcopyright{acmcopyright}\acmConference[ESEM '22]{ACM / IEEE International Symposium on Empirical Software Engineering and Measurement (ESEM)}{September 19--23, 2022}{Helsinki, Finland}
\acmBooktitle{ACM / IEEE International Symposium on Empirical Software Engineering and Measurement (ESEM) (ESEM '22), September 19--23, 2022, Helsinki, Finland}
\acmPrice{15.00}
\acmDOI{10.1145/3544902.3546243}
\acmISBN{978-1-4503-9427-7/22/09}

\usepackage{changepage}
\usepackage{framed}
\makeatletter
 {\par\unskip\endMakeFramed}
\makeatother

\definecolor{formalshade}{rgb}{0.93,0.93,0.93}

\definecolor{darkblue}{rgb}{0.2, 0.2, 0.2}

\newenvironment{formal}{%
  \def\FrameCommand{%
    \hspace{1pt}%
    {\color{darkblue}\vrule width 2pt}%
    {\color{formalshade}\vrule width 4pt}%
    \colorbox{formalshade}%
  }%
  \MakeFramed{\advance\hsize-\width\FrameRestore}%
  \noindent\hspace{-1pt}% disable indenting first paragraph
  \begin{adjustwidth}{}{7pt}%
}
{%
  \end{adjustwidth}\endMakeFramed%
}

\begin{document}

\title{Identifying Source Code File Experts}

\author{Otávio Cury}
\affiliation{%
  \institution{Federal University of Piauí}
  \city{Teresina}
  \country{Brazil}
}
\email{otaviocury@ufpi.edu.br}

\author{Guilherme Avelino}
\affiliation{%
  \institution{Federal University of Piauí}
  \city{Teresina}
  \country{Brazil}
}
\email{gaa@ufpi.edu.br}

\author{Pedro Santos Neto}
\affiliation{%
  \institution{Federal University of Piauí}
  \city{Teresina}
  \country{Brazil}
}
\email{pasn@ufpi.edu.br}

\author{Ricardo Britto}
\affiliation{%
  \institution{Blekinge Institute of Technology}
  \city{Karlskrona}
  \country{Sweden}
}
\email{rbr@bth.se}

\author{Marco Túlio Valente}
\affiliation{%
  \institution{Federal University of Minas Gerais}
  \city{Belo Horizonte}
  \country{Brazil}
}
\email{mtov@dcc.ufmg.br}

\renewcommand{\shortauthors}{Cury, et al.}

\begin{abstract}
  \textbf{Background:} 
  In software development, the identification of source code file experts is an important task. Identifying these experts helps to improve software maintenance and evolution activities, such as developing new features, code reviews, and bug fixes. 
  Although some studies have proposed repository-mining techniques to automatically identify source code experts,  there are still gaps in this area that can be explored.
  For example, investigating new variables related to source code knowledge and applying machine learning aiming to improve the performance of techniques to identify source code experts.
  %The identification of source code file experts is an important task since it can be used to improve software maintenance activities, such as, code review and bug fixes. In this context, it is necessary to create an effective technique for the automatic identification of these experts.
  %There are studies that propose repository-mining techniques to help identifying file experts, but there are still gaps in this area that can be explored. In previous research, it was identified that information such as the \textit{Recency of Modification} and the \textit{File's Size} influence the knowledge that developers have on the file. 
  \textbf{Aim:} 
  The goal of this study is to investigate opportunities to improve the performance of existing techniques to recommend source code files experts.
  %The goal of this study is to analyze the relation of these and other   variables from the development history of a repository with developers' source code knowledge, and compare different techniques for identifying source code file experts. 
  \textbf{Method:} 
  We built an oracle by collecting data from the development history and surveying developers of 113 software projects. Then, we use this oracle to: (i)  analyze the correlation between measures extracted from the development history and the developers’ source code knowledge and (ii) investigate the use of machine learning classifiers by evaluating their performance in identifying source code files experts.
  %We used the development history data from 113 projects in 6 different programming languages. From this analysis, we decided to use machine learning classifiers to guide the identification of file experts. We compare the performances of these classifiers with state-of-the-art techniques, using data from public and private projects. 
  \textbf{Results:} 
  \textit{First Authorship} and \textit{Recency of Modification} are the variables with the highest positive and negative correlations with source code knowledge, respectively. 
  Machine learning classifiers outperformed the linear techniques (F-Measure = 71\% to 73\%) in the public dataset, but this advantage is not clear in the private dataset, with F-Measure ranging from 55\% to 68\% for the linear techniques and 58\% to 67\% for ML techniques.
  \textbf{Conclusion:} Overall, the linear techniques and the machine learning classifiers achieved similar performance, particularly if we analyze F-Measure.  However, machine learning classifiers usually get higher precision while linear techniques obtained the highest recall values. Therefore,  the choice of the best technique depends on the user’s tolerance to false positives and false negatives.
  
  %\textit{First Authorship} and \textit{Recency of Modification} were the variables with the highest positive and negative correlations with source code knowledge respectively. The machine learning classifiers reached the best \textit{F-Measure} ranging from 71\% to 73\%, and the best \textit{Precision} ranging from 71\% to 73\% in both types of projects. \textbf{Conclusion:} Although machine learning classifiers are more accurate in identifying source code experts, we recommend choosing the most appropriate technique according to the user's tolerance for false positives and false negatives.
\end{abstract}




\maketitle

\section{Introduction}
Source code changes are fundamental activities during software evolution \cite{rajlich2014software}. These changes are made in many development-related activities. Such activities require efficient management of the development team. However, this management becomes particularly complicated in large and geographically distributed projects, where project managers need as much information as possible about their development team to coordinate the project activities \cite{herbsleb1999splitting}. In this context, {\bf knowing who has expertise in which parts of the source code is a very useful information}, especially in a context where remote work is growing fast and face-to-face interactions have been reduced \cite{ralph2005pandemic}.

Information on developers' expertise is valuable in various scenarios in software development. For example, it can be used in tasks assignment, such as to identify which experienced developer can help newcomers in implementing changes \cite{kagdi2008can} or who is most suitable for bug fixing \cite{anvik2006should}. Additionally, this information helps to identify the concentration of knowledge in parts of the code \cite{ferreira2017comparison, avelino2016novel}, i.e., situation that poses high risks to the future of the project.

However, due to the large amount of change-related information that developers and managers deal with every day \cite{fritz2014degree}, it is challenging to keep track of who is familiar with each project file. To help with this task, it is possible to rely on information available in Version Control Systems (VCS), wherein a large part of the developer-file iterations are logged. By using such information, several techniques were developed to automate the identification of experts in source code files \cite{fritz2014degree, mcdonald2000expertise, mockus2002expertise, minto2007recommending, da2015niche}.

Some research has been conducted to address the file expert identification problem. For example, in the work \cite{avelino2018can}, the authors compared the performance of three techniques for identifying file experts. They identified {\bf an opportunity for improving the performance of existing techniques by adding information on file size and recency of modifications}. In this paper, we explore this opportunity by first analyzing the correlation between twelve measures extracted from the development history and the developers' source code knowledge. Following, we investigate the use of machine learning classifiers by evaluating their performance in identifying source code file experts on a large dataset composed of public and industrial software systems (including two projects from Ericsson). Particularly, we seek to answer the following research questions: \\

\begin{itemize}
\item \textit{(\textbf{RQ1}) How do repository-based metrics correlate with developer's knowledge?}
\\
\textit{\underline{Motivation:}} There are several works in the literature that use different repository-based metrics to infer the knowledge of developers in source code files. However, we did not identify studies that correlated these variables with knowledge. By answering this question, we seek to understand how these variables are related to knowledge in source code, which can guide the creation of models that estimate knowledge and help to identify source code experts.
\\

\item \textit{(\textbf{RQ2}) How do machine learning classifiers compare with traditional techniques for identifying source code experts?}
\\
\textit{\underline{Motivation:}} Due to the vastly successful application of machine learning classifiers in the software engineering literature, we believe that the application of machine learning classifiers can improve the performance in identifying experts achieved by other techniques in previous works.
\end{itemize}

The main contributions of this paper are twofold: 

\begin{enumerate}
    \item A correlation analysis between variables extracted from version control systems and developers' source code knowledge.
    \item A comparative study on the performances of machine learning classifiers and three well-known techniques for identifying source code experts.
\end{enumerate}

The remainder of this paper is organized as follows: Section \ref{sec:related_work} presents related work. Section \ref{sec:study} describes the procedure adopted to select the target subjects of the study, the compared techniques, and how we evaluate their performance. Sections \ref{sec:results} and \ref{sec:discussion} present the results of the comparison of the techniques and discuss the results, respectively. Section \ref{sec:threats} lists threats to the validity of our results. Finally, Section \ref{sec:conclusion} concludes by presenting our key findings.

\section{Related Work} \label{sec:related_work}

We identified two main goals on research related to the identification of code experts: propose new techniques for the identification of source code experts and compare existing techniques. This section covers both types of works. Section \ref{rel_works:models} presents works that propose techniques to infer developers expertise on source code artifacts and Section \ref{rel_works:evaluation} describes works that compare existing techniques.

\subsection{Research that Proposes New Techniques} \label{rel_works:models}

MacDonald and Ackerman \cite{mcdonald2000expertise} use a heuristic called \textit{Line 10 rule} that prioritizes the developer who last changed a module in solving problems. Following the same premise, Hossen et al. \cite{hossen2014amalgamating} presented an approach called \textit{iMacPro} that identifies experts associated with a change request based on who last changed certain files. Other works count the number of changes made on source code elements \cite{hattori2009mining, hattori2010syde, mockus2002expertise, bird2011don, canfora2012going}. There are also studies that use information from files present in development branches to identify experts who perform merge operations involving these files \cite{costa2016tipmerge, costa2019recommending}. Other models, such as the one proposed by Sülün, Tüzün and Dogrusöz \cite{sulun2019reviewer}, use the number of commits in the artifact of interest and in related artifacts for the calculation of knowledge, in order to recommend code reviewers. In summary, these studies are based mainly on information about changes such as the number of commits and who made the last change to identify expertise. However, based on past works \cite{kruger2018you, avelino2018can}, we suspect that these variables alone are not enough. For this reason, in this study, we analyze more variables and their relationship with developers' knowledge.

Other studies try to model the knowledge flow in the history of the source code. The \textit{Degree of Knowledge} (DOK) model proposed by Fritz et al. \cite{fritz2014degree} uses the information related to the \textit{degree of authorship} (DOA) that the developer has with the code artifact, and the number of interactions (selections and edits) that the developer had with the artifact, named the \textit{degree of interest} (DOI). However, the calculation of the DOI requires the use of special plugins in the development environment, which makes its usage impractical in a large study as the one we present in this paper. Regarding the differences for the models studied in this work, DOK does not deal with recency directly and does not consider the size of the file when estimating knowledge. These two variables were pointed out as important factors in the calculation of knowledge in previous works \cite{kruger2018you, avelino2018can}.

Other techniques model the impact of time on the knowledge that developers have with source code artifacts. Silva et al. \cite{da2015niche} presented a model that computes the developer's expertise in an entire (atomic) artifact, and also in its subparts (internal classes and methods), based on the number of changes made by a developer. The expertise analysis can be done using time windows that divide the history of an artifact into subsets of commits. Other approaches that consider the recency of changes appear in studies focused on the recommendation of developers for the resolution of change requests. Kagdi et al. \cite{kagdi2012assigning} proposed an approach that locates source code files relevant for a given change request and identifies experts in those files using the \textit{xFinder} \cite{kagdi2008can, kagdi2009can} approach, which prioritizes developers who made most commits in a given file. Tüzün and Dogrusöz \cite{sulun2021rstrace+}, extend a previous work \cite{sulun2019reviewer}, by adding information on modification recency for the calculation of knowledge, aiming to recommend code reviewers. On one hand, these studies consider some measure of recency for identifying experts in source code files. On the other hand, they did not present an in-depth and large analysis that shows how the variables used are suitable for this identification.

In comparison to the data source used to extract knowledge information in source code, in this work we focus only on data contained in version control systems. Some works use other sources such as: number of interactions with a file \cite{fritz2007does}, code reviews\cite{thongtanunam2016revisiting, jabrayilzade2022bus}, numbers of meeting related to commits \cite{jabrayilzade2022bus}. While these are valid data sources, they depend on specific tools, such as plugins installed in the development environment, the use of company-specific tools, and development culture. Due to the universality of version control systems in current software development \cite{zolkifli2018version}, its use as a data source becomes easier in practice.

Regarding the use of machine learning, Montandon and colleagues \cite{montandon2019identifying} investigated the performance of supervised and unsupervised classifiers in identifying experts in three open-source libraries. Even though we followed a similar process for data collection and analysis, our work has a distinct purpose. We rely on classifiers for identifying experts at the level of source code files, while Montandon target the identification of experts in the use of libraries and frameworks, therefore using different variables than the ones used in this work.  Other examples of machine learning applications in the context of developer expertise target the bug assignment problem \cite{sajedi2020guidelines}, which is also a distinct problem than the one investigated here. In summary, {\bf we have not identified studies that investigate the performance of machine learning in the classification of file experts based on VCS information}, such as our key goal in this work.

\subsection{Comparison of Existing Techniques} \label{rel_works:evaluation}

Krüger and colleagues \cite{kruger2018you} analyzed the impact of forgetfulness of the developer about the code, using data from ten open-source repositories. They studied whether the forgetting curve described by Ebbinghaus \cite{ebbinghaus1885gedachtnis} can be applied in the context of software development, and which variables influence the developer's familiarity with source code. They analyzed variables such as number of commits, changes made by other developers, percentage of code written by a developer in the current version of the file, and the behavior of tracking changes made by other developers.

Other works used techniques and models to identify expertise. Avelino and colleagues compared the performance of \textit{Commits}, \textit{Blame}, and \textit{Degree-of-Authorship} (DOA) techniques in identifying source code file maintainers \cite{avelino2018can}. A survey similar to the one presented in this paper was made to create a dataset with data from eight open-source repositories and two private ones. The results showed that all three techniques have similar performance in identifying source code maintainers. However, the results also pointed out the importance of considering the recency of the modifications and the file size as a possible strategy to improve these techniques.  

There are also papers that studied other types of expertise. For example, Hannebaur et al. \cite{hannebauer2016automatically} compared the performance of eight algorithms to recommend code reviewers. Out of these algorithms, six are based on expertise by modification and two are based on review expertise. The six algorithms based on expertise by modification are \textit{Line 10 Rule} \cite{mcdonald2000expertise}, Number of Changes, \textit{Expertise Recommender} \cite{mcdonald2000expertise}, Code Ownership \cite{girba2005developers}, \textit{Expertise Cloud} \cite{alonso2008expertise}, and DOA. The algorithms based on review expertise are \textit{File Path Similarity} (FPS) \cite{thongtanunam2014improving}, and a model proposed by the authors, called \textit{Weighted Review Count} (WRC). They used data from four FLOSS projects: Firefox, AOSP, OpenStack, and Qt. The algorithms based on review expertise performed better than the ones based on modification expertise, and the WRC algorithm achieved the best results.

Other works use information on expertise for the resolution of bug reports. Anvik and Murphy \cite{anvik2007determining} compared two approaches to determine appropriate developers for resolving bug reports. One approach uses data from code repositories to define which developers are experts in the files associated with the bug report using the \textit{Line 10} heuristic. The other approach uses data from bug networks such as the carbon-copy list (cc:), comments, and information from who resolved previous bugs. Development data from the Eclipse\footnote{https://www.eclipse.org/eclipse/} platform was used. The authors concluded that the best approach depends on what users are looking for regarding precision and recall.

The work presented in this paper can be distinguished from the works described before in three key aspects: purpose, compared techniques, and scope. In relation to purpose, two works have the same objectives as ours: Avelino et al. \cite{avelino2018can} and Anvik and Murphy \cite{anvik2007determining} (but they do not use the same techniques, particularly machine learning classifiers).

Regarding the compared techniques, Avelino et al. \cite{avelino2018can}, Krüger et al. \cite{kruger2018you}, and Hannebauer et al. \cite{hannebauer2016automatically} used the baseline techniques selected in this work. However, we also investigate the performance of machine learning models. Finally, regarding the scope of the studies, none of them used data from a similar number of repositories as in our study.

\section{Research Design}
\label{sec:study}

To achieve the objectives defined in this study, it is required a ground truth with data on source code experts. We build this ground truth by extracting data from open source and industrial projects. This ground truth is composed of the developers' knowledge in source code files and variables (or metrics) computed from the projects' development history.

\subsection{Target Subjects} \label{sub:opensource}
We selected open source repositories from the GitHub platform\footnote{https://github.com/}. To select these repositories, we adopted a similar procedure to other studies that investigate GitHub data \cite{avelino2016novel, yamashita2015revisiting, ray2014large}. First, for each of the six most popular programming languages on GitHub (Java, Python, Ruby, JavaScript, PHP, and C++)\footnote{The six most popular programming languages in 2019 https://octoverse.github.com/\#top-languages} we selected the 50 most popular repositories as indicated by their number of stars. This measure is widely used by researchers in the selection of GitHub repositories \cite{avelino2016novel, ray2014large, padhye2014study, hilton2016usage, mazinanian2017understanding, jiang2017understanding, nielebock2019programmers, rigger2018analysis, castro2018analysis}, and perceived by developers as a reliable proxy of popularity \cite{borges2018s}. Then, after cloning the repositories,  we performed a repository filtering step based on three metrics: number of commits, number of files, and number of developers. As was done in a previous work \cite{avelino2016novel}, for each language, we removed repositories in the first quartile of the distribution of each metric, resulting in the intersection of the remaining sets. In other words, repositories with few commits, files, and developers were removed. Table \ref{table:q1} shows the first quartile of each of the three metrics for each programming language.

\begin{table}[t]
\centering
\caption{First quartiles of filtering metrics, for each language}
\label{table:q1}
\begin{tabular}{lrrr}
\hline
\multicolumn{1}{c}{\textbf{Language}} & \multicolumn{1}{c}{\textbf{Commits}} & \multicolumn{1}{c}{\textbf{Files}} & \multicolumn{1}{c}{\textbf{Developers}} \\ \hline
Python                                & 510.75                               & 87.50                              & 45.25                                   \\
Java                                  & 829.25                               & 318.00                                & 39.25                                   \\
PHP                                   & 823.50                               & 89.00                                 & 97.50                                   \\
Ruby                                  & 1,650.25                             & 152.75                             & 198.25                                  \\
C++                                   & 2,010.25                             & 706.00                                & 113.50                                  \\
JavaScript                            & 1,455.75                             & 113.25                             & 129.25                                  \\ \hline
\end{tabular}
\end{table}

Additionally, we discarded repositories whose development history suggests that most of the software was developed outside of GitHub by removing repositories where more than half of its files were added in a few commits. As few commits, we considered the outliers of the distribution of the number of files added in each commit of a repository; we discarded repositories if most of their files (>50\%) were added by the set of outliers commits, following the procedure done by Avelino and others \cite{avelino2016novel}.

The resulting dataset is composed of {\bf 111 popular GitHub repositories} distributed over the six most popular programming languages, which have a relevant number of developers, files, and commits. Table \ref{table:repos} summarizes the characteristics of these 111 repositories.

\begin{table}[t]
\centering
\caption{Target open source repositories.}
\label{table:repos}
\begin{tabular}{@{}lrrrr@{}}
\toprule
\multicolumn{1}{c}{\textbf{Language}} & \multicolumn{1}{c}{\textbf{Repos}} & \multicolumn{1}{c}{\textbf{Devs}} & \multicolumn{1}{c}{\textbf{Commits}} & \multicolumn{1}{c}{\textbf{Files}} \\ \midrule
Python                                & 25                                 & 18,936                            & 192,587                              & 39,154                             \\
Java                                  & 17                                 & 5,733                             & 138,473                              & 62,429                             \\
PHP                                   & 16                                 & 9,802                             & 144,092                              & 29,902                             \\
Ruby                                  & 25                                 & 38,036                            & 605,546                              & 88,869                             \\
C++                                   & 14                                 & 11,467                            & 350,345                              & 72,991                             \\
JavaScript                            & 14                                 & 10,319                            & 109,541                              & 21,477                             \\ \midrule
Total:                                & 111                                & 94,293                            & 1,540,584                            & 314,822                            \\ \bottomrule
\end{tabular}
\end{table}

We also used data from {\bf two industrial projects from Ericsson}\footnote{https://www.ericsson.com/en}. Both projects were developed in the Java programming language. The development history of Project \#1 has 74,078 commits, 17,329 files, and 513 contributors. On the other hand, Project \#2 has 26,678 commits, 15,930 files, and 262 contributors. Therefore, the two industrial projects are also relevant according to the criteria applied to select the open-source repositories.

\subsection{Ground Truth Construction} \label{sec:ground_truth}

\noindent \textbf{Extracting Development History}: after downloading the selected repositories, we started the process of extracting their development history data. This data was extracted by collecting the commits from the \textit{master/default-branch} of each repository, as described below.

First, we ran \tcode{git log --no-merge --find-renames} command to extract data from the commits logs of each repository. This command returns all commits that have no more than one parent (no-merge) and it automatically handles possible file renames.\footnote{https://git-scm.com/docs/git-log/1.5.6} From each commit, three pieces of information were extracted: (1) changed files; (2) name and email of the  commit's author (developer who performed the change); (3) type of the change: \textit{addition}, \textit{modification}, or \textit{rename}.

After that, we discarded files that do not contain source code (e.g., images and documentation), third-party libraries, and files that are not part of one of the six programming languages considered in the study. To identify and discard these files we rely on the Linguist tool \footnote{https://github.com/github/linguist/blob/master/lib/linguist/languages.yml}.

Finally,  we handled developer aliases by following the same procedure adopted in other works \cite{avelino2018can, avelino2016novel, avelino2017assessing, Avelino2019a}. 
Aliases arise when a developer is associated with more than one pair \textit{(name-dev, email)} on Git. For the purposes of this work, it is important to unify these contributors. This unification was performed in two stages. First, users who shared the same email, but with different names, were unified. Additionally, users with different emails, but with similar names were grouped. This similarity was computed using the \textit{Levenshtein} distance \cite{navarro2001guided}, using a maximum modification threshold of 30\% on the number of letters.\\

\noindent \textbf{Generating Survey Sample: }We used information extracted from development histories to create sample pairs \textit{(developer, file)} for each repository. These samples are necessary to elaborate the survey applied in this study. They were created by performing the following steps for each repository:

\begin{enumerate}
  \item We randomly selected a file and retrieved the list of developers who touched (created or modified) it.
  
  \item We discarded the file if at least one of these developers reached the maximal limit of files we plan to send to a developer (\textit{file\_limit}), asking about his/her expertise on the file. Otherwise, we added the file to the list of each developer.
  
  \item We repeated steps 1-2 until there are no more files to be verified.
\end{enumerate}

By discarding files in each one of the developers already reached her \textit{file\_limit}, step 2 warrants that a file will be added to the sample only if it is possible to add all developers who modified it. This step aims to maximize the possibility of obtaining answers from all developers who touched a file.
We established a \textit{file limit} of five, seeking to not discourage developers from responding to the survey, which can happen according to guidelines in the literature \cite{kasunic2005designing, linaker2015guidelines}.

At the end of this step, a list of pairs \textit{(developer, file)} was created for each repository. For open-source repositories, 20,564 pairs were generated, with 7,803 developers, 2.64 files per developer on average. For the two private repositories, 394 pairs were generated, with 92 developers, and 4.34 files per developer.
\\

\noindent \textbf{Sending the Survey}: We conducted the survey by sending individual e-mails to each developer on the generated sample. The developers were invited to evaluate their knowledge in each of the files on their list, by using a scale from 1 (one) to 5 (five), where
\textit{(1) means you have no knowledge about the file's code; (3) means you would need to perform some investigations before reproducing the code; and (5) means you are an expert on this code}

For the open-source repositories, 7,803 emails were sent, and 501 responses were received, representing a response rate of 7\%. For the private repositories, 92 emails were sent, 38 responses were received, representing a rate of 41\%. From these responses, two datasets were created: one with 1,024 developer-file pairs coming from the answers of the open-source repositories, and a second dataset with 163 pairs extracted from the answers of the two private repositories. In the remainder of this paper, these two datasets will be named public and private datasets.
\\

\noindent \textbf{Processing the Answers:} \label{processing_answers}
We process the answers by classifying each pair \textit{(developer, file)} into one of two disjoint sets: \textit{declared experts} ($O_{m}$), and \textit{declared non-experts} ($O_ {\overline{m}}$). A \textit{declared expert} is a developer who claims to have knowledge of more than 3 (three) in a file; otherwise, he/she is a \textit{declared non-expert}.

\begin{figure}[t]
  {\includegraphics[width=0.4\textwidth]{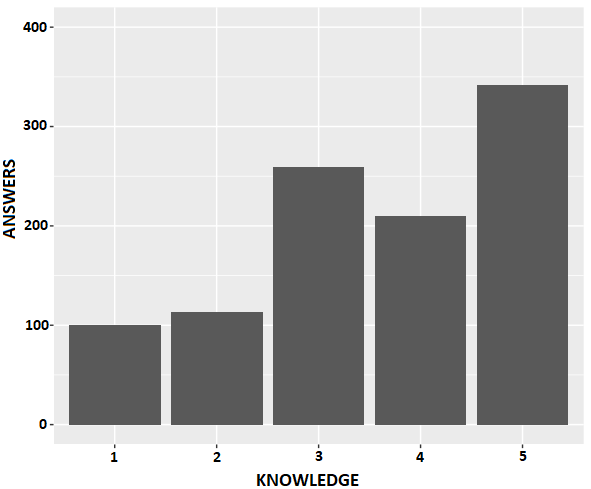}\caption{Responses distribution in the Public dataset.}\label{fig:public}}
 \end{figure}

\begin{figure}[t]  
  {\includegraphics[width=0.4\textwidth]{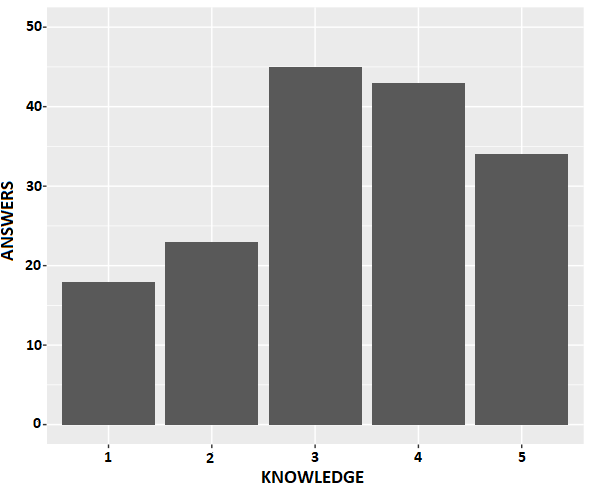}\caption{Responses distribution in the Private dataset}\label{fig:private}}
\end{figure}

Figure \ref{fig:public} and \ref{fig:private} shows the distribution of responses in the public and private datasets respectively. \textbf{The public dataset is composed of 54\% of \textit{declared experts} and 46\% of \textit{declared non-experts}, and the private dataset is composed of 47\% of \textit{declared experts} and 53\% of \textit{declared non-experts}}. As we can see, the proportion of declared expert and declared non-expert sets are close in both datasets, and this class balance is an important property in classifier training \cite{weiss2001effect}.


\subsection{Development Variables}

\noindent \textbf{Extracting Variables:} \label{sec:vars}
We selected 12 variables (or features) that can be extracted from the development history. These variables and their meanings are shown in Table \ref{table:variables}. Subsets of these variables are explored in other studies. Table \ref{table:extracted} lists studies that analyzed these variables in the context of inferring source code knowledge. As shown, \textit{NumCommits} is the most explored variable.

\begin{table*}[t]
\centering
\caption{Variables extracted from the development history. The  variables description are given considering a developer \textit{d} and a file \textit{f} in its last version.}
\label{table:variables}
\begin{tabular}{ll}
\hline
\multicolumn{1}{c}{\textbf{Variable}} & \multicolumn{1}{c}{\textbf{Meaning}}                                                                                                       \\ \hline
Adds                                  & Number of lines added by a developer \textit{d} on a file \textit{f}                                                                  \\
Dels                                  & \begin{tabular}[c]{@{}l@{}}Number of lines deleted by a developer \textit{d} on a file \textit{f}\end{tabular}                    \\
Mods                                  & \begin{tabular}[c]{@{}l@{}}Number of lines modified by a developer \textit{d} on a file \textit{f}\end{tabular}                    \\
Conds                                 & \begin{tabular}[c]{@{}l@{}}Number of conditional statements added by a developer \textit{d} on a file \textit{f}\end{tabular}     \\
Amount                                & \begin{tabular}[c]{@{}l@{}}Sum of number of added and deleted lines of a developer \textit{d} on a file\end{tabular}   \\
FA                                    & \begin{tabular}[c]{@{}l@{}}Binary variable that indicates whether a developer \textit{d} added the file \textit{f} to\\  the repository\end{tabular}          \\
Blame                                 & \begin{tabular}[c]{@{}l@{}}Number of lines authored by a developer \textit{d} that are in a file \textit{f}\end{tabular}                \\
NumCommits                            & \begin{tabular}[c]{@{}l@{}}Number of commits made by a developer \textit{d} on a file \textit{f}\end{tabular}                     \\
NumDays                               & Number of days since the last commit of a developer \textit{d} on a file \textit{f}                                                                      \\
NumModDevs                            & \begin{tabular}[c]{@{}l@{}}Number of developers that committed on the file \textit{f} since the last \\ commit of a developer \textit{d}\end{tabular}    \\
Size                                  & Number of lines of code (LOC) of the file \textit{f}                                                                                                \\
AvgDaysCommits                        & \begin{tabular}[c]{@{}l@{}}Average number of days between the commits of a developer \textit{d} \\ on a file \textit{f}\end{tabular} \\ \hline
\end{tabular}
\end{table*}

The variables \textit{Adds}, \textit{Dels}, \textit{Mods}, and \textit{Conds} are extracted using the command \tcode{git diff} \footnote{https://git-scm.com/docs/git-diff}. This command returns the lines added and removed between two versions of a file. In this work, a modification is defined as a set of removed lines followed by a set of added lines of the same size \cite{werneyMaster, werney}. We use \textit{Levenshtein} distance \cite{navarro2001guided} to identify which pairs \textit{(remove, add)} are modifications between two versions of the files. The algorithm takes the removed and added line as input, and returns a value that represents the number of characters that must be modified to transform the removed line into the added line. In this work, a change is a modification if the returned value is less than a certain percentage (threshold) of the size of the removed line. A threshold of 40\% was used, as suggested by Canfora et al. \cite{canfora2007identifying}. \\

\begin{table}[t]
\centering
\caption{Extracted variables and related studies}
\label{table:extracted}
\begin{tabular}{@{}ll@{}}
\toprule
\multicolumn{1}{c}{\textbf{Variable}} & \multicolumn{1}{c}{\textbf{Also used in these studies}}                                                                                                                                                                                                                                                                                      \\ \midrule
Adds, Dels                            & \cite{werneyMaster, werney, de2018analysis, mockus2002expertise, girba2005developers, hattori2009mining}                                                                                                                                                                                               \\
Mods, Conds, Amount                   & \cite{werneyMaster, werney, de2018analysis}                                                                                                                                                                                                                                                            \\
Blame                                 & \cite{girba2005developers, rahman2011ownership, kruger2018you, avelino2018can, hannebauer2016automatically}                                                                                                                                                                                            \\
FA                                    & \cite{fritz2014degree, fritz2007does}                                                                                                                                                                                                                                                                  \\
NumCommits                            & \cite{werneyMaster, werney, de2018analysis, kruger2018you, avelino2018can, fritz2014degree, montandon2019identifying, oliveira2019well, mockus2002expertise, mcdonald2000expertise, alonso2008expertise, constantinou2016identifying} \\ \cite{hannebauer2016automatically, da2015niche, minto2007recommending, sun2017enhancing, kagdi2012assigning, kagdi2009can} \\
NumDays                               & \cite{werneyMaster, werney, de2018analysis, kruger2018you, avelino2018can, sun2017enhancing}                                                                                                                                                                                                                             \\
NumModDevs                               & \cite{werneyMaster, werney, de2018analysis, fritz2014degree}                                                                                                                                                                                                                                           \\
Size                                  & \cite{avelino2018can, oliveira2019well}                                                                                                                                                                                                                                                                \\
AvgDaysCommits                        & \cite{montandon2019identifying}                                                                                                                                                                                                                                                                        \\ \bottomrule
\end{tabular}
\end{table}

\subsection{Compared techniques} \label{sec:compared_tec}

This section describes the techniques used in this study to identify code experts.
\\

\noindent \textbf{Number of Commits:} \label{tec:commits}
This technique counts the number of commits as a measure of knowledge that a developer has in a given source code file. 
The idea behind this technique is that a developer gains knowledge in a file by creating or modifying it. 
In other words, more commits are interpreted as more knowledge. This technique is widely used in the context of developer expertise, as shown in Table \ref{table:extracted}. It is applied individually \cite{avelino2018can, da2015niche, hannebauer2016automatically}, as well as combined with other techniques \cite{fritz2014degree, sun2017enhancing, kagdi2012assigning}.   
In this study, we use the number of commits individually as an expert identification technique. 
In this paper, we refer to this technique as \textit{NumCommits}.
\\

\noindent \textbf{Blame:} \label{tec:blame}
This technique infers the knowledge that a developer has in a file by counting the number of lines added by him and present in the last version of the file. Blame-like tools such as \tcode{git-blame}\footnote{https://git-scm.com/docs/git-blame} command are used to identify the authors of each file line. As in other works \cite{avelino2018can, rahman2011ownership, girba2005developers, kruger2018you, hannebauer2016automatically}, we use the percentage of lines associated with a developer as a measure of knowledge.
\\

\noindent \textbf{Degree of Authorship (DOA):} \label{tec:doa}
Fritz et al. \cite{fritz2014degree} proposed that the knowledge of a developer on a source code file depends on factors such as the file's authorship, his/her number of contributions, and number of changes made by other developers. The authors combined these variables into a linear model called \textit{Degree of Authorship} (DOA). The weights associated with each variable in this linear model were defined through an empirical study based on data from the development of two systems. The DOA value that a developer \textit{d} has in the  version \textit{v} of a \textit{f} file is calculated by the following equation:

\begin{equation} \label{eq:doa}
    \textbf{DOA(d, f(v))} = 3.293 + 1.098 * FA + 0.164 * DL - 0.321 * ln(1 + AC)
\end{equation}

where,

\begin{itemize}
  \item $FA$: 1 if developer \textit{d} is the creator of the file \textit{f}, 0 otherwise.
  \item $DL$: is the number of changes made by developer \textit{d} until version \textit{v} of file \textit{f}.
  \item $AC$: is the number of changes made by other developers in the file \textit{f} up to version \textit{v}.
\end{itemize}

\noindent \textbf{Machine Learning Classifiers:} \label{tec:machine_classifiers}
We also investigate the applicability of machine learning algorithms for identifying source code experts. This proposal comes down to a binary classification. From the dataset described in Section \ref{sec:ground_truth}, we labeled each pair \textit{(developer, file)} in non-expert (knowledge 1-3, as their answers in the survey) and experts (knowledge 4-5). We choose the development variables (Table \ref{table:extracted}) \textit{Adds}, \textit{FA}, \textit{Size}, and \textit{NumDays} as features to train the machine learning models. The rationale for these choices is given in Section \ref{sec:results}, where we present the results of the \textit{RQ1}. According to the values of these features, machine learning models can classify a developer as an expert or not from a source code file.

Five well-known machine learning classifiers are evaluated: Random Forest \cite{liaw2002classification}, Support-Vector Machines (SVM) \cite{weston1998multi}, K-Nearest Neighbor (KNN) \cite{peterson2009k}, Gradient Boosting Machine \cite{friedman2001greedy}, and Logistic Regression \cite{hosmer2013applied}. We use \textit{10-fold Cross-Validation} to evaluate the performance of the machine learning classifiers. 10-fold Cross-Validation consists of partitioning the data set in ten complementary subsets, and the use of each of these subsets for model training and the rest for model validation. In the end, the fitness measures are combined to obtain a more accurate estimate of the model's performance \cite{berrar2019cross}. We use \textit{F-Measure} as a performance measure. To find the best settings for these classifiers, we perform a grid search \cite{claesen2015hyperparameter} to adjust the hyper-parameters and find the best settings for each classifier.
\\

\subsection{Linear Techniques Evaluation}

To evaluate \textit{NumCommits}, \textit{Blame}, and \textit{DOA} in the identification of software experts, we adopt a similar procedure as the one followed in a previous related work \cite{avelino2018can}.
First, we normalize the technique values. For this purpose, we set as $1$ the expertise of a developer \textit{d} in file \textit{f} if \textit{d} has the highest value for a given technique in \textit{f}; otherwise, we set a proportional value. For example, suppose that for a given file \textit{f} developers \textit{d1}, \textit{d2}, and \textit{d3} have a \textit{Blame} value of 10, 15, and 20. Their normalized values for blame regarding the file \textit{f} are as follows: \textit{expertise(d1, f)} = 10/20 = 0.5, \textit{expertise(d2, f)} = 15/20 = 0.75, and \textit{expertise(d3, f)} = 20/20 = 1. We apply this normalization process to all techniques.

After the normalization, a developer \textit{d} is classified as an expert of a file \textit{f} if its \textit{expertise (d, f)} is higher or equal than a threshold \textit{k}; otherwise he/she is considered a \textit{non-expert}. In the example of the previous paragraph, considering a threshold \textit{k} = 0.7, developers \textit{d2} and \textit{d3} would be considered experts of the \textit{f} file accordingly with the \textit{Blame} technique.
\\

\noindent \textbf{Evaluating Performance:} \label{linear_evaluation}
We always compared the performance of the technique in their best settings. For the linear techniques, this requires setting the classification threshold \textit{k} that maximizes the correct identification of file experts.

To this purpose, we first compute the performance of each technique by adopting 11 different thresholds, i.e., by varying the threshold \textit{k} from 0 to 1, under 0.1 steps. At the end of this process, the best performance of the linear techniques is obtained together with their associated thresholds \textit{k}. For each threshold, we use \textit{10-fold Cross-Validation} to compute the \textit{F-Measure} (F1-Score) of the techniques by applying the following equations:

\begin{equation}
Precision(k) = 
\begin{cases}
    \frac{|(d, f)\, \in \, O_{m} \ |\, \textit{expertise(d, f)} \, > \, \textit{k}|}{|\, \textit{expertise(d, f)} \, > \, \textit{k} \,|},& \text{if } k = 0\\
    \frac{|(d, f) \, \in \, O_{m} \ |\, \textit{expertise(d, f)} \, \geq \, \textit{k}|}{|\textit{expertise(d, f)} \, \geq \, \textit{k} \,|},              & \text{otherwise}
\end{cases}
\end{equation}

\begin{equation}
Recall(k) = 
\begin{cases}
\frac{|(d, f) \, \in \, O_{m} \ |\ \textit{expertise(d, f)} \, > \, \textit{k}|}{|O_{m}|},& \text{if } k = 0\\
\frac{|(d, f) \, \in \, O_{m} \ |\ \textit{expertise(d, f)} \, \geq \, \textit{k}|}{|O_{m}|},& \text{otherwise}
\end{cases}
\end{equation}

\begin{equation}
F-Measure(k) = 2*\frac{Precision(k)*Recall(k)}{Precision(k)+Recall(k)}
\end{equation}

\noindent where
$O_{m}$ is the set of declared experts, as we inferred from the survey responses (Section \ref{processing_answers}). 
\\

\noindent \textbf{Thresholds Calibration:} \label{thresholds_calibration}
 
\begin{figure}[t]
  \centering
  {\includegraphics[width=0.4\textwidth]{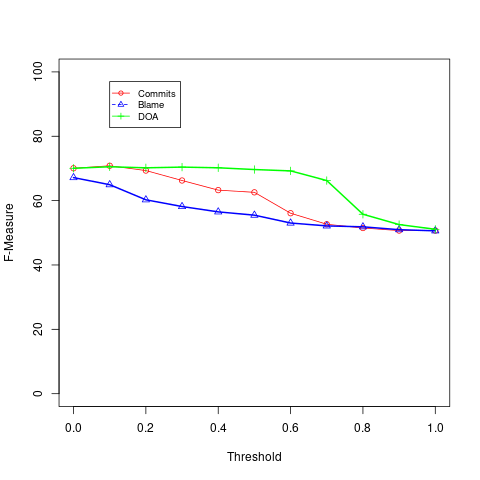}\caption{Performance at analyzed thresholds using public data}\label{fig:resulPublic}}
 \end{figure}

\begin{figure}[t]
  {\includegraphics[width=0.4\textwidth]{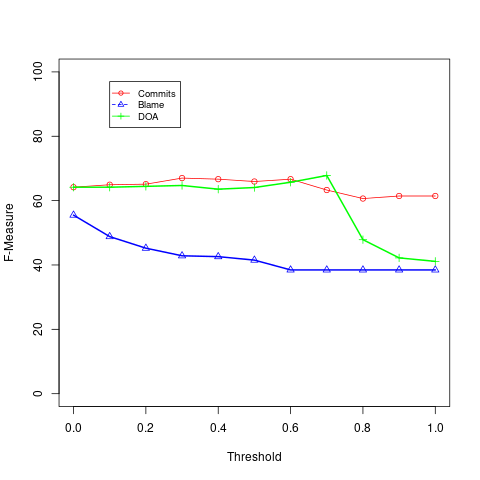}\caption{Performance at analyzed thresholds using private data}\label{fig:resulPrivate}}
\end{figure}

Figures \ref{fig:resulPublic} and \ref{fig:resulPrivate} show the \textit{F-Measure} results for each linear technique at each analyzed threshold, under the public and private datasets, respectively. 


As we can see, \textit{Blame} reaches its best performance when we use the lowest possible threshold ($\mathit{k = 0}$), in both public and private data. In other words, by adopting $\mathit{k =0}$, Blame reaches its best performance when it is used to classify as an expert any developer who has at least one line of code in the last version of the file. Similarly, \textit{NumCommits} achieves the best performance by adopting the thresholds \textit{k} = 0.1 and \textit{k} = 0.3, in the public and private datasets, respectively. The low thresholds indicate that both techniques perform better by relying on minimal changes to consider developers as file experts. On the other hand, \textit{DOA} has it best performance with a threshold \textit{k} = 0.1. on public data; and \textit{k} = 0.7, on private data.

\section{Results} \label{sec:results}

This section presents the results of the two research questions proposed in this study. First, we present the correlation analysis between the extracted variables and developers' knowledge. Next, we present the performance of the techniques in our two datasets.

\subsection{How do repository-based metrics correlate with developer's knowledge? \textit{(RQ1)}}

We analyze the correlations between the development variables and the developers' knowledge, as obtained in the survey. In the remainder of this paper, the survey responses will be represented by a variable named \textit{Knowledge}. We answer \textit{RQ1} and, consequently, identify which variables could be part of a prediction knowledge model.

Table \ref{table:cor} shows the directions and intensities of the correlations between the extracted variables and the \textit{Knowledge} variable, by applying Spearman's \textit{rho}, since it is suitable for data sets that do not follow the normal distribution. We consider results to be statistically significant when \textit{p} < 0.05. Even though \textit{NumCommits} is the most used variable for inferring file's knowledge in the literature, it does not show the strongest positive correlation with the \textit{Knowledge} variable. The variable with the highest positive correlation is \textit{FA}, closely followed by \textit{Adds}. This suggests that a more fine-grained measure of changes like \textit{Adds} can be more important than \textit{NumCommits} to compose a knowledge model. The variable with the lowest correlation with \textit{Knowledge} is  \textit{Mods}. Finally, the variable that shows the highest negative correlation is \textit{NumDays}, followed by \textit{NumModDevs} and \textit{Size}. These results reinforce the importance of recency for inferring the knowledge that a developer has about a source code file.

\begin{table}[t]
\centering
\caption{Correlation of extracted variables with \textit{Knowledge}}
\begin{tabular}{@{}lrr@{}}
\toprule
\multicolumn{1}{c}{\textbf{Variable}} & \multicolumn{1}{c}{\textbf{Corr. with Knowledge}} & \multicolumn{1}{c}{\textbf{p-value}} \\ \midrule
NumDays                               & - 0.24                                            & \textless{}0.001                     \\
Size                                  & - 0.21                                            & \textless{}0.001                     \\
NumModDevs                            & - 0.21                                            & \textless{}0.001                     \\
Mods                                  & 0.01                                              & 0.515                                \\
Dels                                  & 0.02                                              & 0.369                                \\
Cond                                  & 0.19                                              & \textless{}0.001                     \\
NumCommits                            & 0.20                                              & \textless{}0.001                     \\
AvgDaysCommits                        & 0.21                                              & \textless{}0.001                     \\
Amount                                & 0.28                                              & \textless{}0.001                     \\
Blame                                 & 0.29                                              & \textless{}0.001                     \\
Adds                                  & 0.31                                              & \textless{}0.001                     \\
FA                                    & 0.33                                              & \textless{}0.001                     \\ \bottomrule
\end{tabular}
\label{table:cor}
\end{table}

\begin{figure}[t]
  {\includegraphics[width=0.4\textwidth]{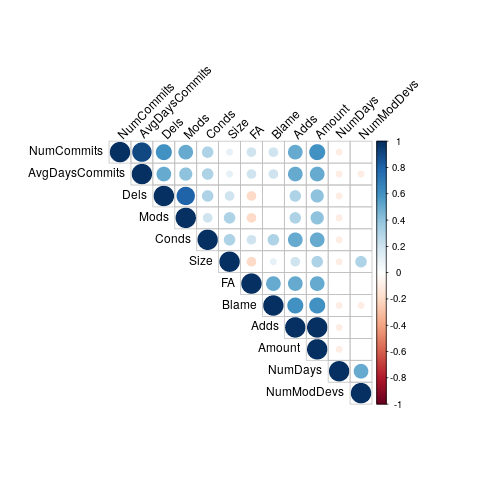}\caption{Correlation between dependent variables}\label{fig:corrs}}
\end{figure}

We also analyze the correlation between the extracted variables using Spearman's \textit{rho} \cite{schober2018correlation}. These correlations are represented in Figure 5, where colors close to 1 and -1 represent higher positive and negative correlations, respectively.
This analysis is relevant because variables with a certain level of correlation can lead to inaccurate models \cite{yu2003feature}. 

As expected,  \textit{NumCommits} has a strong correlation with  \textit{Adds}, \textit{Dels}, \textit{Mods}, \textit{Amount}, and \textit{AvgDaysCommits} (\textit{rho} $\geq$ 0.5).
Therefore, any one of these five variables can be used to measure the number of changes made by a developer throughout the history of a file. The variables \textit{NumModDevs} and \textit{NumDays} also show a moderate positive correlation (\textit{rho} $\geq$ 0.5) with each other.

\begin{formal}
\textbf{Summary of RQ1:} \textit{First Authorship} (FA) has the highest positive correlation with knowledge in source code files and \textit{Recency of Modification} has the highest negative correlation. Therefore, this result suggests that file creators tend to have high knowledge on the files they created; however, this knowledge decreases with time.
\end{formal}

\noindent \textbf{Variable Selection Rationale:}
In a previous research, Avelino and colleagues described how the variables \textit{file size} and \textit{recency} influence developer's knowledge on source code files~\cite{avelino2018can}. Thus, models that take this information into account tend to be more accurate. This previous finding is reinforced by the results in Table \ref{table:cor}, where these variables achieved the highest negative correlation with \textit{Knowledge}. Therefore, we include the variables \textit{Size} and \textit{NumDays} in the proposed machine learning  models. Considering the variables \textit{Adds}, \textit{Amount}, and \textit{Blame},  we choose the variable \textit{Adds}, as it has the highest positive correlation with \textit{Knowledge} among the three. No more than one of the three variables is chosen because these variables are conceptually related, as commented in Section \ref{sec:vars}.

The variable \textit{NumCommits} is not included in the models. It can be viewed as just a macro way of accounting for changes made by a developer to a source code file, and the variable \textit{Adds} has already been chosen, since it has a higher positive correlation with \textit{Knowledge}. In addition, \textit{Adds} has a moderate correlation with \textit{NumCommits} ($ rho \geq 0.5 $, Figure \ref{fig:corrs}). For a similar reason,  \textit{AvgDaysCommits} is not included in the model.

We also add the variable \textit{FA}, which has the highest  positive correlation with \textit{Knowledge} among all variables. The variable \textit{NumModDevs} is not included due to its moderate correlation  with \textit{NumDays} ($ rho \geq 0.5 $).

\subsection{How do machine learning classifiers compare with traditional techniques in identifying source code experts? \textit{(RQ2)}}

After identifying and selecting the variables with the highest correlations with source code knowledge, we train the machine learning models using them as features. After that, we compared the performance of these models with those of linear models in the identification of experts.

Table \ref{table:result} presents the performance of linear techniques and machine learning classifiers, in the two analyzed scenarios.
Regarding the public dataset (Table \ref{table:result} - \textbf{Public}), \textit{DOA} and \textit{NumCommits} had the best F-Measure (70\%), followed by \textit{Blame}, with a  F-Measure of 67\%. Regarding recall, the technique with the best result was \textit{DOA} (97\%). The technique with the lowest recall in the same scenario is \textit{Blame} (83\%). Finally, the technique with the highest precision is \textit{NumCommits} (60\%), while \textit{DOA} has the lowest precision (55\%).

\begin{table*}[t]
\caption{Performance of linear and machine learning techniques}
\label{table:result}
\begin{tabular}{@{}lrrrrrr@{}}
\toprule
\textbf{}                    & \multicolumn{3}{c}{\textbf{Public}}                                                                                   & \multicolumn{3}{c}{\textbf{Private}}                                                                                           \\ \midrule
                             & \multicolumn{1}{c}{\textbf{Precision}} & \multicolumn{1}{c}{\textbf{Recall}} & \multicolumn{1}{c}{\textbf{F-Measure}} & \multicolumn{1}{c}{\textbf{Precision}} & \multicolumn{1}{c}{\textbf{Recall}} & \multicolumn{1}{c}{\textbf{F-Measure}} \\
\textbf{DOA}                 & 0.55                                   & 0.97                                & 0.70                                   & 0.61                                   & 0.77                                & 0.68                                   \\
\textbf{NumCommits}          & 0.60                                   & 0.87                                & 0.70                                   & 0.55                                   & 0.86                                & 0.67                                   \\
\textbf{Blame}               & 0.56                                   & 0.83                                & 0.67                                   & 0.50                                   & 0.62                                & 0.55                                   \\ \midrule
\textbf{Random Forest}       & 0.73                                   & 0.74                                & 0.73                                   & 0.59                                   & 0.62                                & 0.60                                   \\
\textbf{SVM}                 & 0.71                                   & 0.75                                & 0.73                                   & 0.63                                   & 0.60                                & 0.61                                   \\
\textbf{KNN}                 & 0.71                                   & 0.71                                & 0.71                                   & 0.70                                   & 0.69                                & 0.67                                   \\
\textbf{GBM}                 & 0.73                                   & 0.73                                & 0.73                                   & 0.65                                   & 0.58                                & 0.60                                   \\
\textbf{Logistic Regression} & 0.69                                   & 0.75                                & 0.72                                   & 0.70                                   & 0.51                                & 0.58                                   \\ \bottomrule
\end{tabular}
\end{table*}

When using the private dataset (Table \ref{table:result} - \textbf{Private}), \textit{DOA} had the highest value for F-Measure (68\%), closely followed by \textit{NumCommits} (67\%). As in the other scenario,  \textit{Blame}  had the worst F-Measure (55\%). The best Recall was from \textit{NumCommits} (86\%), with \textit{Blame} presenting the lowest recall (62\%). Regarding precision, \textit{DOA} achieved the best result (61\%), and \textit{Blame} the lowest one (50\%).

When we consider the scenario using the public dataset (Table \ref{table:result} - \textbf{Public}), \textit{Random Forest}, \textit{SVM}, and \textit{GBM} had the best F-Measure (73\%), with the other two classifiers reaching close values. The classifiers with the highest recall were \textit{SVM} and \textit{Logistic Regression} (75\%), and \textit{KNN} obtained the lowest result (71\%). In terms of  precision, \textit{GBM} and \textit{Random Forest} have the highest values (73\%). The lowest precision is from \textit{Logistic Regression} (69\%).

Regarding the private dataset (Table \ref{table:result} - \textbf{Private}), the classifier with the highest F-Measure was \textit{KNN} (67\%) and \textit{Logistic Regression} had the lowest one (58\%). The best recall was also obtained by  \textit{KNN} (69\%) and \textit{Logistic Regression} had the lowest one (51\%). Finally, the best precision was achieved by \textit{Logistic Regression} and \textit{KNN} (70\%). \textit{Random Forest} had the lowest precision (59\%).

\begin{formal}
\textbf{Summary of RQ2:} In the public dataset, machine learning classifiers outperformed the linear techniques (\textit{F-Measure} = 0.71 to 0.73). In the private dataset, this advantage is not clear, with F-Measure ranging from 0.55 to 0.68 for the linear techniques and 0.58 to 0.67 for ML techniques.
\end{formal}

\section{Discussion} \label{sec:discussion}

In this section, we start by highlighting our key findings:

\begin{enumerate}

\item The linear techniques (DOA, Commits, and Blame) tend to have low precision, but high recall in the two analyzed datasets (public and industrial projects). In other words, they tend to positively classify developers as experts, but at the cost of a significant number of false positives.

\item Regarding the linear techniques, DOA has the best performance, in both datasets. 

\item The ML classifiers have a very distinct performance in the investigated datasets. However, we must highlight the size difference of both datasets: there are 1,024 datapoints in the public dataset and only 163 in the private one. This difference can have a major impact on algorithms that include a training phase. In other words, ML techniques tend to improve their performance as the size of the training set increases, up to a certain point of stability, a behavior called the \textit{Learning Curve} \cite{meek2002learning}. For this reason, we argue the public dataset results are more representative of the performances of such techniques.

\item Considering the public dataset, the best ML classifiers (in terms of F-Measure) are  \textit{Random Forest}, \textit{SVM}, and \textit{Gradient Boosting Machines}. On the other hand, in the private dataset, \textit{K Nearest Neighbors} reached the best performance. 

\item Even though there are linear techniques with similar \textit{F-Measure}, the machine learning classifiers show a better balance between \textit{Precision} and \textit{Recall}. 

\end{enumerate}

However, overall, we acknowledge that the linear techniques and the machine learning classifiers achieved similar performance, particularly if we analyze F-Measure. For this reason, {\bf the choice of the best technique depends on the user's tolerance to false positives and false negatives.} Particularly, we envision two practical application scenarios:

\begin{itemize}
    \item When the application requires finding few or even just one expert, such as when performing a merge operation \cite{costa2016tipmerge} or onboarding a new team member \cite{kagdi2009can}, machine learning classifiers are more recommended as they are more precise. 

\item When the application demands more experts, such as when we need to evolve or maintain a more complex feature \cite{hossen2014amalgamating}, it is more plausible to tolerate some false positives. In such cases, linear techniques are recommended.

\end{itemize}

\section{Threats to Validity} \label{sec:threats}

This section describes the threats to the validity of this study based on four categories: construct, internal, conclusion, and external validity \cite{wohlin2012experimentation}.
\\

\noindent \textbf{Construct Validity:} Some of the developers who participated in the survey may have inflated their self-assessment knowledge, due to some fear regarding the commercial use of this information. To tackle this problem, we clarified in the survey that their answers would be used only for academic purposes. Regarding the risk of misinterpretation of the knowledge scale (1 to 5), we provided a guide to the scores meaning. However, even the concept of an expert is subjective, which remains a threat. Another threat is the definition of expert according to the survey responses. Developers with knowledge above three were considered file experts.  We claim this is a more conservative division of the knowledge scale, already applied on a similar study \cite{avelino2018can}.
\\

\noindent \textbf{Internal Validity:} There are other factors that may have an influence on the knowledge that a developer has in a source code file, which might not have been taken into account in this study. For example, knowledge can be acquired in activities that do not require commits in VCS. However, in this work we limit ourselves to identify experts using authorship information that can be extracted from version control systems, which are tools widely used in software development, making the evaluated techniques applicable to most projects. Nevertheless, there are other variables that can be extracted from the information contained in code repositories, which can also play an important role in the process of identifying file experts. Among them, we can mention the iteration with files in the same module and the complexity of the changed code. However, we perceive these variables as more complex, more difficult to compute, and dependent on the programming language and design of the target systems. For this reason, we focused on variables that are less dependent on the particularities of the software under analysis.
\\

\noindent \textbf{Conclusion Validity:} Regarding the statistical tests, Spearman's \textit{rho} is used to analyze the correlation between the extracted variables. This coefficient was chosen because it does not require a normal distribution or a linear relationship between the variables. However, there is no wide consensus on the interpretation of the correlation values returned by the test. However, we relied on interpretations that are also used in studies from different areas \cite{schober2018correlation, overholser2008biostatistics}. We use 10-fold cross-validation to assess the performance of the machine learning models. We emphasize the classifier results may vary to some degree according to the number of folds chosen in the cross-validation. However, in general, five or 10-folds are recommended as good parameters \cite{hastie2009elements}.
\\

\noindent \textbf{External Validity:} Some decisions were taken to maximize the ability to generalize the study results. For open-source systems, we choose six popular programming languages. Therefore, by not limiting the projects to a single programming language, we intended to reduce the impact that certain languages may have on the developers' familiarity assessment, and consequently in our results. In total, data from 113 systems are used. The small amount of data from private projects makes it difficult to generalize the results for this type of project, but these results can be generalized to projects with characteristics similar to those analyzed.

\section{Conclusion} \label{sec:conclusion}

In this paper, we investigated the use of predictive models for the identification of source code file experts based on information available in version control systems. From a large dataset composed of public and industrial projects, we identified the variables that are most related to source code knowledge (RQ1). \textit{First Authorship} and \textit{Number of Lines Added} shown highest positive correlations with knowledge, while \textit{Recency} and \textit{File Size} have the highest negative correlation. Using the best variables in this analysis, machine learning models were evaluated for the identification of experts (RQ2). The performance of these models was compared with other techniques previously studied in the literature. As a result, the ML algorithms achieved the best F-Measure and precision among the compared techniques, with highlights for \textit{Random Forest}, \textit{Support Vector Machines}, and \textit{Gradient Boosting Machines}.

Our findings can support the design and implementation of tools that seek to automate the process of identifying file experts in software projects. We also provided insights on the variables that are most relevant to the derivation of new knowledge models. The models evaluated in this paper can be part of future investigations and research in repository analysis. In addition, future research can assess the performance of the studied techniques in other industrial contexts.

As a final note, we provide a replication package with the results of our analysis as well as the survey’s answers, repositories data, and scripts used in the paper. The replication package is available at https://doi.org/10.5281/zenodo.6757349.

\section{Competing interests}

The authors declare that they have no known competing financial interests or personal relationships that could have appeared to influence the work reported in this paper. 

\begin{acks}

We would like to thank the 501 developers (Ericsson and Open Source projects) who took their time to answer the survey and provided essential information to our investigation.
Additionally, we also thank CAPES, CNPq, FAPEMIG, and UFPI for supporting this work.

\end{acks}

\bibliographystyle{ACM-Reference-Format}
\bibliography{sample-base}

\end{document}